# Authentication Schemes Using Braid Groups

*Sunder Lal and Atul Chaturvedi*
(Department of Mathematics
Institute of Basic Science
Khandari, Agra-282002(UP)-India)
sunder_lal2,@rediffmail.com
atulibs@gmail.com

**Abstract:** In this paper we proposed two identification schemes based on the root problem. The proposed schemes are secure against passive attacks assuming that the root problem (RP) is hard in braid groups.

## 1. Introduction

The idea of using the braid group as a platform for cryptosystems was introduced in [2]. In recent years have emerged several proposals for secure cryptographical schemes using noncommutative groups, in particular Artin's braid groups [1, 2, 4, 5, and 7]. Braid groups are more complicated than Abelian groups, and are not too complicated to be worked with. These two characteristics make braid group a natural choice. In fact, the Conjugacy Problem (CP) and the Root Problem (RP) in braid groups are algorithmically difficult, and it consequently provide one-way functions. We use it to propose two identification scheme based on Root Problem over a braid group.

It is well known that an *identification scheme* is an important and useful cryptographic tool. The identification scheme is an interactive protocol where a prover, **P**, tries to convince a verifier, **V**, of his identity. Only **P** knows the secret value corresponding to his public one, and the use of this secret value allows **P** to convince **V** of its identity.

The rest of the paper is organized as follows: We present a brief introduction of braid groups in section 2. In section 3, we define identification schemes. In section 4, we present our identification schemes, and we give a proof of security and zero-knowledge for our schemes. The paper ends with conclusion.

## 2. Braid Groups

Emil Artin [3] in 1925 defined $B_n$, the braid group of index n, using following generators and relations: Consider the

**generators** $\sigma_1, \sigma_2, ..., \sigma_{n-1}$, where $\sigma_i$ represents the braid in which the $(i+1)^{st}$ string crosses over the $i^{th}$ string while all other strings remain uncrossed.

The **definining relations** are

(1)   $\sigma_i \sigma_j = \sigma_j \sigma_i \text{ for } |i-j| > 1$

(2)   $\sigma_i \sigma_j \sigma_i = \sigma_j \sigma_i \sigma_j \text{ for } |i-j| = 1$.

The reader is referred to any textbook about braids for a geometrical interpretation of each element of the group $B_n$ by an n-strand braid in the usual sense. The braid $\Delta = (\sigma_1 \sigma_2 \ldots \sigma_{n-1})(\sigma_1 \sigma_2 \ldots \sigma_{n-2}) \ldots (\sigma_1 \sigma_2)(\sigma_1)$ is called the *fundamental braid*. $\Delta$ nearly commutes with any braid b. In fact $\Delta b = \tau(b)\Delta$, where $\tau: B_n \to B_n : \tau(\sigma_i) = \sigma_{n-i}$ is an automorphism. Since $\tau^2$ is the identity map, $\Delta^2$ truly commutes with any braid. A subword of the fundamental braid $\Delta$ is called a *permutation braid* and the set of all permutation braids is in one-to-one correspondence with the set $\sum_n$ of permutations on $\{0, 1, \ldots, n-1\}$. For example, $\Delta$ is the permutation sending $i$ to $n-i$. The word length of a permutation n-braid is $\leq \frac{n(n-1)}{2}$. The *descant set* $D(\pi)$ of a permutation $\pi$ is defined by $D(\pi) = \{i | \pi(i) > \pi(i+1)\}$. Any braid b can be written uniquely as $b = \Delta^u \pi_1 \pi_2 \ldots \pi_l$ where u is an integer, $\pi_i$ are permutation braids such that $\pi_i \neq \Delta$ and $D(\pi_{i+1}) \subset D(\pi_i^{-1})$. This unique decomposition of a braid b is called a *left canonical form*.

*All the braids in this paper are supposed to be in the left-canonical form.* For example, for $a, b \in B_n$, ab means the left-canonical form of ab and so it is hard to guess its factors a or b from ab.

If b is a non-trivial and $e \geq 2$ is an integer, then $b^e$ is never identity. In other words, the braid groups are torsion-free.

For given $y \in B_n$ and $e \geq 2$ finding x such that $y = x^e$ is called **root problem (RP)**.

It is proved in [**8**] that RP is decidable but it is computationally infeasible when braids of a sufficient size are considered.

## 3. Identification Schemes

An identification scheme or entity authentication protocol, allows one party to gain assurances that the identity of another is as declared. It is used to prevent impersonation.

It is an interactive protocol which involves a *prover* or claimant **P** and a *verifier* **V**. In general, **P** tries to convince the verifier **V** of his identity. The verifier is presented with, or presumes beforehand, the purported identity of the prover. The goal is to corroborate that the identity of the prover is indeed **P**. Only **P** knows the secret value corresponding to his public one, and it is the proper use of this secret value which allows **P** to convince **V** of the identity of **P**.

The objectives of an identification protocol include the following.
1. In the case of honest parties **P** and **V**, **P** is able to successfully authenticate himself to **V**, i.e., **V** will complete the protocol having accepted **P**'s identity.
2. (Transferability) **V** cannot reuse an identification exchange with **P** so as to successfully impersonate **P** to a third party **A**.
3. (Impersonation) The probability is negligible that any party **C** distinct from **P**, carrying out the protocol and playing the role of **P**, can cause **V** to accept **P**'s identity.
4. The previous points hold even if: a polynomial large number of previous authentication between **P** and **V** have been observed; the adversary **A** has participated in previous protocol executions with either or both **P** and **V**; and multiple instances of the protocol, possibly initiated by **C**, may be run simultaneously.

## 4. The Proposed Schemes

Here we propose two identification schemes. For both the schemes, the initial setup is as follows:

Let $B_n$ be a braid group where RP is infeasible. As mentioned earlier, all the braids in $B_n$ are assumed to be in the left canonical form. Thus for a, b in $B_n$, it is hard to guess a or b from ab. We assume that n is even, and denote by $LB_n$ (resp. $UB_n$) the subgroup of $B_n$ generated by $\sigma_1, ..., \sigma_{\frac{n}{2}-1}$, i.e., braids where the n/2 lower strands only are braided ( resp. in the subgroup generated by $\sigma_{\frac{n}{2}+1}, ..., \sigma_{n-1}$). We know that every element in $LB_n$ commutes with every element in $UB_n$. We also take H as a fixed collision-free hash function on $B_n$.

**Scheme I : Key Generation**

  **P** generates private and public keys as follows:
  (a) **P** choose two integers $r \geq 2$, and $s \geq 2$;
  (b) **P** chooses $a \in LB_n$, and $b \in UB_n$ such that RP for a, b is hard enough;
  (c) **P** computes $X = a^r b^s$;
  (d) **P**'s public key is (X,r,s); and the secret key is the pair (a, b).

**Authentication**
  (a) **V** choose $c \in UB_n$, $d \in LB_n$, and sends the challenge $Y = c^r d^s$ to **P**.
  (b) **P** sends the response $Z = H(a^r Y b^s)$ to **V**.
  (c) **V** checks $Z = H(c^r X d^s)$.

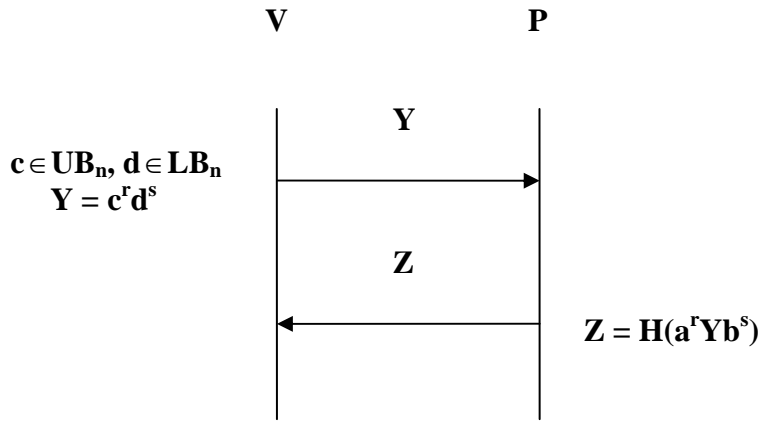

Figure1: Proposed Scheme I

**Security:**
**Completeness:** Assume that, at step 2(b), **P** sent Z'. Then **V** accepts **P**'s key if and only if we have $Z' = H(c^r X d^s)$. The latter relation is equivalent to
$Z' = H(c^r(a^r b^s)d^s)$……………(1)
By hypothesis, $a,d \in LB_n$ while $b,c \in UB_n$, so $ac = ca$ and $bd = db$. Therefore (1) is equivalent to $Z' = H(a^r(c^r d^s)b^s)$, i.e., to $Z' = Z$.

**Soundness:** Assume a cheater **C** is accepted with non-negligible probability. This means that **C** can compute $H(c^r X d^s)$ with non-negligible probability. As H is supposed to be an ideal hash function, this means that **C** can compute a braid z satisfying $H(z) = H(c^r X d^s)$ with non-negligible probability. There are two possibilities: either we have $z = c^r X d^s$, which contradicts the hypothesis that the RP for c and d is hard, or $z \neq c^r X d^s$, which means that **C** and **V** are able to find a collision for H, contradicting the hypothesis that H is collision free.

**Honest-Verifier Zero-Knowledge:** Consider the probabilistic Turing machine defined as follows: it chooses random braids c and d using the same drawing as the honest verifier, and outputs the instances $(c,d,H(c^r X d^s))$. Then, the instance generated by this simulator follows the same probability distribution as the ones generated by the interactive pair (**P,V**).

**Scheme II:** In this scheme the initial setup is the same. However **P** generates his keys as follows:

**Key Generation**
1. **P** generates a sufficiently complicated braid s in $B_n$.
2. **P** Choose two integer $e \geq 2$, and $f \geq 2$.
3. **P** Choose $a \in LB_n$, and computes $X = a^e s a^f$.
4. The public key is Pub = $< n, e, f, s, X >$, and the secret key is $< a >$.

**Authentication**

Protocol consists of k-times challenge-response protocol where k is security parameter as usual identification protocol. The two pass challenge-response protocol is described as follows:
1. **V** chooses b in $UB_n$, and sends the challenge $Y = b^e s b^f$ to **P**.
2. **P** sends the response $Z = H(a^e Y a^f)$ to **V**.
3. **V** accepts **P**'s proof of identity if and only if $Z = H(b^e X b^f)$ and reject otherwise.

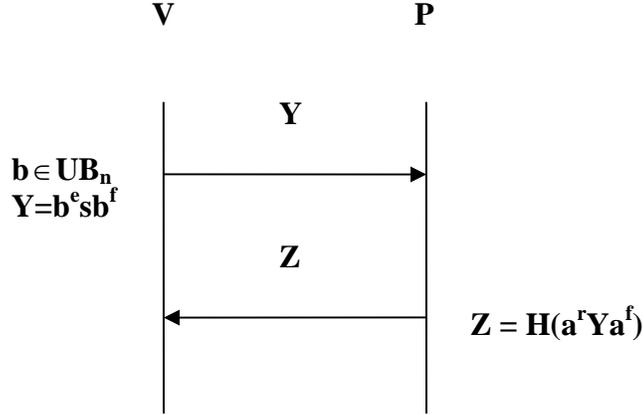

Figure2: Proposed Scheme II

**Security:**
**Completeness:** Assume that, at Phase II (2), **P** sends T. The **V** accepts **P**'s proof if and only if $T = H(b^e X b^f)$, i.e., if and only if $T = H(b^e(a^e s a^f)b^f)$. By hypothesis $a \in LB_n$ while $b \in UB_n$, so $ab = ba$. Therefore $T = H(a^e(b^e s b^f)a^f) = H(a^e Y a^f) = Z$.

**Soundness:** Assume a cheater **C** is accepted with non-negligible probability. This means that **C** can compute $H(b^e X b^f)$ with non-negligible probability. As H is supposed to be an ideal hash function, this means that **C** can compute a braid z satisfying $H(z) = H(b^e X b^f)$ with non-negligible probability. There are two possibilities: either we have $z = b^e X b^f$, which contradicts the hypothesis that the RP for b is hard, or $z \neq b^e X b^f$, which means that **C** and **V** are able to find a collision for H, contradicting the hypothesis that H is collision free.

**Honest-Verifier Zero-Knowledge:** Consider the probabilistic Turing machine defined as follows: it chooses random braids b using the same drawing as the honest verifier, and outputs the instances $(b, H(b^e X b^f))$. Then, the instance generated by this simulator follows the same probability distribution as the ones generated by the interactive pair (**P**, **V**).

**Conclusion:** In this paper, we have proposed some identification schemes specially designed for braid groups. Our schemes are two-pass protocol based on RP and we give security proof of our schemes.